# COMPUTER VISION FOR VOLUNTEER COTTON DETECTION IN A CORN FIELD WITH UAS REMOTE SENSING IMAGERY AND SPOT-SPRAY APPLICATIONS

Pappu Kumar Yadav[1*], J. Alex Thomasson[2], Stephen W. Searcy[1], Robert G. Hardin[1], Ulisses Braga-Neto[3], Sorin C. Popescu[4], Daniel E. Martin[5], Roberto Rodriguez[6], Karem Meza[7], Juan Enciso[1], Jorge Solórzano Diaz[8], Tianyi Wang[9]

*Corresponding Author: Pappu Kumar Yadav (pappuyadav@tamu.edu)

107, Price Hobgood Agricultural Engineering Research Lab

375 Olsen Blvd, College Station, TX 77840

[1] Department of Biological & Agricultural Engineering, Texas A&M University, College Station, TX

[2] Department of Agricultural & Biological Engineering, Mississippi State University, Mississippi State, MS

[3] Department of Electrical & Computer Engineering, Texas A&M University, College Station, TX

[4] Department of Ecology & Conservation Biology, Texas A&M University, College Station, TX

[5] Aerial Application Technology Research, U.S.D.A. Agriculture Research Service, College Station, TX

[6] U.S.D.A.- APHIS PPQ S&T, Mission Lab, Edinburg, TX

[7] Department of Civil & Environmental Engineering, Utah State University, Logan, UT

[8] Texas A&M AgriLife Research & Extension Center, Weslaco, TX

[9] College of Engineering, China Agricultural University, Beijing, China

**Highlights**

- Computer vision algorithm was developed with YOLOv5m for detecting volunteer cotton (VC) plants in cornfields.
- Pixel based bounding box coordinates were converted into geodetic GPS coordinates of detected VC plants.
- Optimal flight path was generated using ant colony optimization algorithm (ACO) for spot-spray UAS.
- Spot-spray UAS based on optimal flight path was simulated using Mission Planner and MAVProxy ground control station software.

**ABSTRACT.** *To control boll weevil (Anthonomus grandis L.) pest re-infestation in cotton fields, the current practices of volunteer cotton (VC) (Gossypium hirsutum L.) plant detection in fields of rotation crops like corn (Zea mays L.) and sorghum (Sorghum bicolor L.) involve manual field scouting at the edges of fields. This leads to many VC plants growing in the middle of fields remain undetected that continue to grow side by side along with corn and sorghum. When they reach pinhead squaring stage (5-6 leaves), they can serve as hosts for the boll weevil pests. Therefore, it is required to detect, locate and then precisely spot-spray them with chemicals. In this paper, we present application of YOLOv5m on radiometrically and gamma corrected low resolution (1.2 Megapixel) multispectral imagery for detecting and locating VC plants growing in the middle of tasseling (VT) growth stage of cornfield. Our results show that, VC plants can be detected with a mean average precision (mAP) of 79% and classification accuracy of 78% on images of size 1207 x 923 pixels at an average inference speed of nearly 47 frames per second (FPS) on NVIDIA Tesla P100 GPU-16GB and 0.4 FPS on NVIDIA Jetson TX2 GPU. We also demonstrate the application of a customized unmanned aircraft systems (UAS) for spot-spray applications based on the developed computer vision (CV) algorithm and how it can be used for near real-time detection and mitigation of VC plants growing in corn fields for efficient management of the boll weevil pests.*

# INTRODUCTION

The cotton boll weevil (*Anthonomus grandis* L.) is an insect pest that has caused more than 23 billion USD in losses to the U.S. cotton industry since migrating from Mexico in the 1890s (Harden, 2018). It continues to be a matter of concern for the U.S. cotton industry, particularly in Texas even though it has been eradicated from most of the U.S. Therefore, there is a continued need for the activities of the Texas Boll Weevil Eradication Foundation (TBWEF), as per the latest report of Sunset Advisory Commission (Roming et al., 2021). The TBWEF has divided the state into 16 eradication zones, in which the Lower Rio Grande Valley (LRGV) is still actively functioning, as it remains the region most prone to boll weevil re-infestation each year due to its tropical climatic conditions (Roming et al., 2021) and close proximity to the Mexico border. In 2019 alone, 46,000 boll weevils were captured by the foundation indicating severity of the problem and a continued need for functioning of the TBWEF.

Cotton (*Gossypium hirsutum* L.) is commonly planted in rotation with crops like corn (*Zea mays* L.) and sorghum (*Sorghum bicolor* L.). In climatic areas like the LRGV, cotton seeds can survive year round, and thus seeds in cotton that might have fallen during harvest in the previous year can grow among corn and sorghum plants in the present year (Wang et al., 2022; Yadav et al.,2022.). Such plants are called volunteer cotton (VC) plants, which essentially act as weeds but are mainly of concern because they can serve as hosts for boll weevils. To minimize the likelihood of boll weevil re-infestation, TBWEF uses pheromone traps to detect boll weevil and pesticides to eliminate them. As part of boll weevil mitigation efforts in the LRGV region, fields with rotation crops are inspected for the presence of VC plants at the edges of fields on a weekly basis. When VC plants are detected, the number of pheromone traps are increased. In addition to inspecting for the presence of VC plants, pheromone traps are also inspected for the presence of boll weevils. If at least one is found at the edge of a field, then the entire field is sprayed with a pesticide, commonly Malathion ($C_{10}H_{19}O_6PS_2$) ULV (FYFANON® ULV AG) at rates between 0.56 and 1.12 kg/ha (FMC Corporation, 2001). Uniform spraying is done because VC plants growing in the middle of corn and sorghum fields remain undetected and so cannot be sprayed individually. Spraying entire fields results in increased management cost, and environmental concerns as well as destroying many beneficial insects.

Uniform spray applications can be avoided if VC plants growing in the middle of corn and sorghum fields are detected and precisely located so that spot-spray capable unmanned aircraft systems (UAS) can be deployed. Detecting VC plants before they reach pinhead square stage and precisely spraying them with herbicides can eliminate the plants and minimize the need for applying Malathion. However, some VC plants may survive due to herbicide tolerance or inaccuracies in detection, in which case Malathion application later in the season could be needed. To detect VC plants either early or late in the growing season, remote sensing with images collected by UAS along with computer vision (CV) algorithms using state-of-the-art convolution neural network (CNN) architectures like Mask R-CNN (He et al., 2017), YOLOv3 (Redmon and Farhadi, 2018; Yadav et al., 2022a, 2022b), YOLOv5 (Jocher et al., 2021a), etc., can be used. Since its release in 2021, YOLOv5 has become popular in CV applications and has been used to detect various objects like apples (Kuznetsova et al., 2021), face masks (Vinay Sharma, 2020), safety helmets (Zhou et al., 2021), etc. Due to its higher detection accuracy and faster inference speed, YOLOv5 was selected for this study as the most viable model for near real-time detection.

In addition to detecting VC plants, the geographic coordinates of the detected plants are needed for precise spray application. Therefore, geotagged UAS-based imagery was used in this study. High quality RGB (red, green, blue) cameras have commonly been used for object detection with YOLOv3 and YOLOv5 (Kuznetsova et al., 2021; Li et al., 2022; Sharma, 2020). In most cases, radiometric correction was not employed. Remote sensing imagery without radiometric correction is susceptible to varying environmental conditions including illumination, atmospheric light scattering, sensor noise, etc. (Hausamann et al., 2005). Images that are not radiometrically corrected have digital numbers (DN) that do not represent actual surface reflectance (Biday and Bhosle, 2012; Mamaghani and Salvaggio, 2019; Minařík et al., 2019). Hence, radiometric correction was conducted in this study.

YOLO series object detection algorithms generate bounding boxes (BB) around the objects of interest present in the images. Locations of these BBs are based on pixel-wise distance from the top-left corner of images (Alexey et al., 2021; Jocher et al., 2021c; Redmon and Farhadi, 2018). Based on the BB coordinates, central coordinates of each BB can be determined. Pixel-wise coordinates are not useful for path planning of spot-spray capable UAS, so a method of converting pixel-wise BB coordinates into geographic coordinates is necessary so they can be used for path planning for spot-spraying detected VC plants.

UAS flight times are limited by battery capacity, so an optimal flight path is necessity to efficiently spray VC plants. Optimal path planning can be conducted based on the travelling salesman problem (TSP), in which the goal is to determine the shortest route for the UAS to spot-spray each of the detected VC plants and then return back to the starting point (Shivgan and Dong, 2020; Sorma et al., 2020). Different algorithms have been tested for this, some of which include the genetic algorithm by Moon et al. (2002) and Shivgan and Dong (2020), ant colony optimization (ACO) by Dorigo et al. (2021), etc. In this study, ACO was used because of simplicity in implementation, feasibility, and faster speed to generate high quality solutions (Dorigo et al., 2021). The determined optimal flight path was tested by simulating the UAS with DroneKit-Software In The Loop (SITL) (3D Robotics, Berkely, California, U.S.A.), MAVProxy (Tridgell and Barker, 2009), and Mission Planner (Ardupilot Development Team and Community). DroneKit-SITL provides an application programming interface (API) to run Python based applications on companion computers of UAS, whereas MAVProxy and Mission Planner are ground control station (GCS) software to control and simulate UAS applications (Qays et al., 2020). The communication between GCS and UAS is done through a binary serial telemetry protocol called MAVLink (Meier, 2009).

The overall goal of this research was to develop a CV algorithm for detecting VC plants in a corn field and use the detected locations for optimal spot-spray applications. The four specific objectives were (i) to develop a computer vision (CV) algorithm with YOLOv5m to detect VC plants in a corn field with radiometrically (reflectance calibrated) and gamma corrected, relatively low-resolution (1.2 Megapixel), multispectral aerial imagery; (ii) to convert pixel-based bounding box coordinates of detected VC plants into geographic coordinates; (iii) to use the detected geographic coordinates of VC plants to generate an optimal flight path with the ACO algorithm; and (iv) to simulate spot-spray UAS with DroneKit-SITL, MAVProxy and Mission Planner GCS software based on the generated optimal flight path.

# MATERIALS AND METHODS

## EXPERIMENT SITE

This study was conducted at a corn field (fig. 1; 96°25'45.9"W, 30°32'07.4"N) of roughly 5.9 hectares (14.6 acres) at the Texas A&M AgriLife Research farm near College Station, Texas. The majority of the soil in the experimental plot is Weswood silty clay loam, followed by Yahola fine sandy loam and Belk clay (USDA-Natural Resources Conservation Service, 2020). The corn plants in the field were in the second leaf (V2) vegetative state when 90 cotton seeds each of two varieties (Phytogen 340 W3FE, CORTEVA agriscience, Wilmington, Delaware; and Deltapine 1646 B2XF, Bayer AG, Leverkusen, North Rhine-Westphalia, Germany) were planted at randomized locations among the corn plants to mimic the presence of VC plants in the field. Some were planted in line with corn plants while others were planted in the furrow middles.

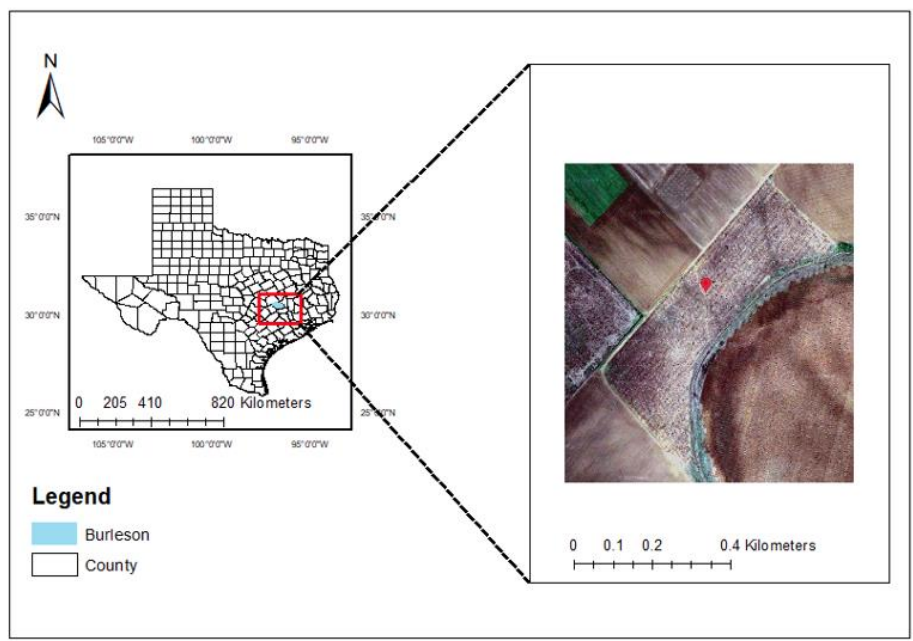

**Figure 1. Experiment field located at Texas A&M University farm near College Station, TX in Burleson county where cotton plants were planted in the middle of corn field to mimic the presence of volunteer cotton plants.**

**IMAGE DATA ACQUISITION**

A five-band (Blue, Green, Red, Near Infrared, RedEdge) multispectral camera : RedEdge-MX (AgEagle Aerial Systems Inc., d/b/a MicaSense, Wichita, Kansas; fig. 2) was mounted on a customized UAS (fig. 2) for collecting aerial imagery of the corn field with cotton plants when the corn plants had reached tassel vegetation stage (VT) (fig. 3). The images were acquired at this growth stage of corn plants so that VC detection accuracy could be tested when the corn plants were relatively larger and taller than the VC plants. This was done as a part of study in which accuracies at different growth stages of corn plants were compared. The UAS (Hylio AG-110; Hylio Inc., Richmond, Texas) was originally designed for broadcast spray applications. Data were collected on May 14, 2021, between 11:00 a.m. and 2:00 p.m. central daylight-saving time (CDT) at an altitude of nearly 4.6 meter (15 feet) above ground level. This resulted in an approximate ground sampling distance of 0.34 cm/pixel. Since the customized UAS was not designed for aerial surveying, there was no software interface available to capture images based on overlap settings. Therefore, images were captured on a timer-based settings enabling an image to capture every second. This resulted in many unused and distorted images due to unavailability of overlap settings and vibrations caused due to the design aspect of the sprayer UAS. The UAS was flown at a speed of 2 meter per second.

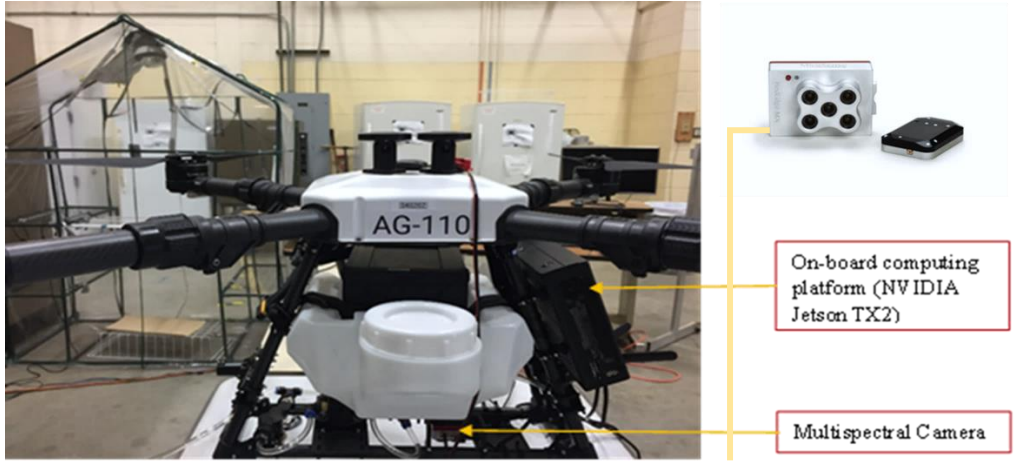

**Figure 2. A customized sprayer UAS with RedEdge-MX multispectral camera for capturing aerial imagery and NVIDIA Jetson TX2 computing platform.**

An onboard computing platform, the Jetson TX2 (NVIDIA Corporation, Santa Clara, CA, U.S.A.) development board that consists of a Pascal graphics processing unit (GPU), was mounted on the customized UAS with the intent of near real-time detection of VC plants in corn fields (fig. 2). The Pascal GPU is low-cost, fast, and widely used as an embedded artificial intelligence computing device. It consists of 256 NVIDIA Compute Unified Device Architecture (CUDA; NVIDIA, Santa Clara, CA, U.S.A.) cores with 8 GB of RAM and 32 GB of storage capacity (NVIDIA, 2022).

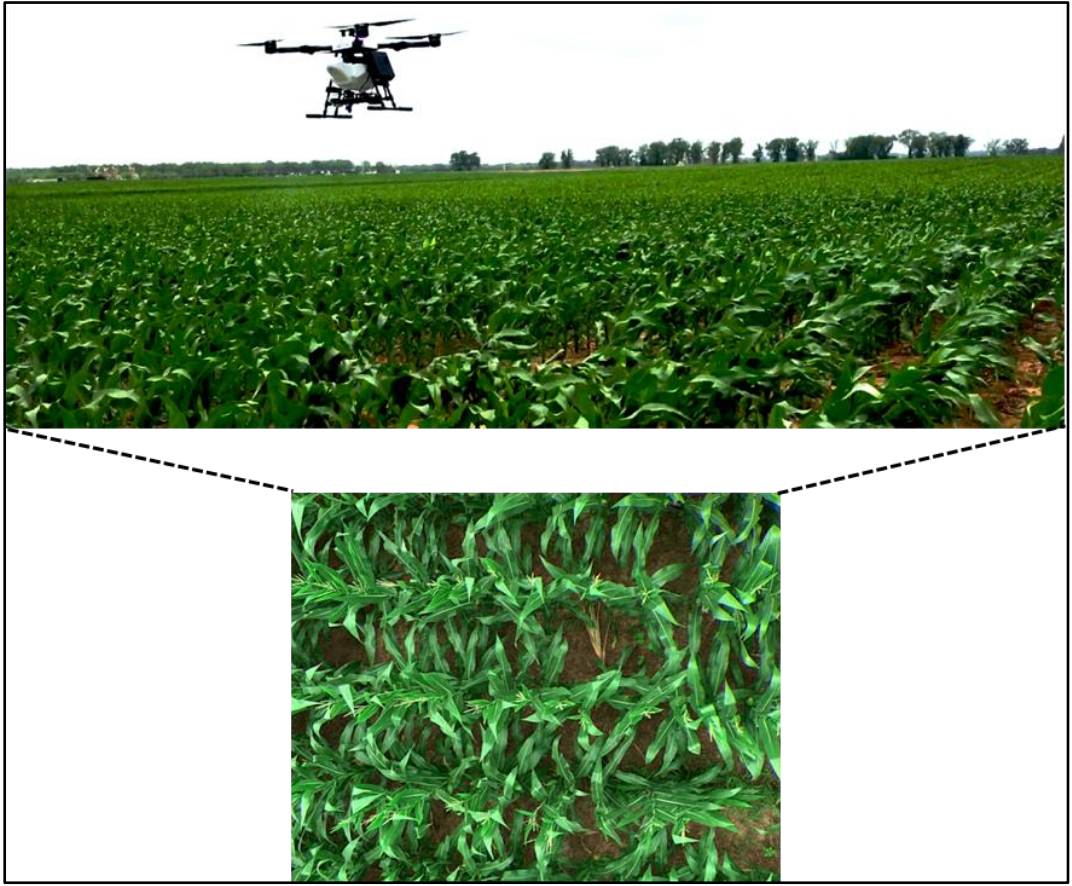

**Figure 3. RGB (Red, Green Blue) composite image showing a section of experimental plot where corn at vegetative tassel state (VT) can be seen.**

## MANUFACTURER RECOMMENDED CORRECTIONS (RADIOMETRIC AND GAMMA)

The individual band images collected by RedEdge-MX camera were corrected based on manufacturer's recommendations using the reflectance panel (fig. 4) images taken on the day of data collection just before the flight.

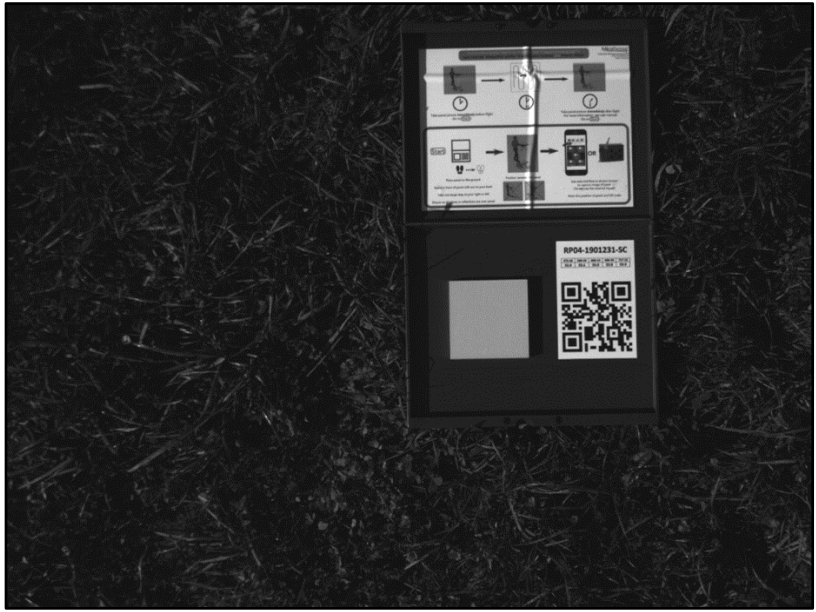

**Figure 4. Reflectance panel image with blue band sensor of RedEdge-MX camera taken on the day of flight.**

According to the manufacturer's factory calibration (in which absolute reflectance values were plotted along wavelengths ranging from 400 nm to 850 nm), the reflectance values corresponding to blue, green, red, NIR and RedEdge were 0.60, 0.61, 0.61, 0.60 and 0.56 respectively for the panel that was used in the study. As per the manufacturer's process, area of the Lambertian panel from reflectance panel image is extracted and its radiance value is converted to the scale of reflectance value which is then applied to the whole image to convert into reflectance images. Source code for both radiometric and gamma corrections were used from the GitHub (GitHub, Inc., San Francisco, CA, U.S.A.) repository of MicaSense (MicaSense Incorporated, 2022). The original source code was modified based on the reflectance values for the panel used as well as for generating RGB aligned images for each captured images from the field data collection as seen in figure 3. Python version 3.8.12 (Python Software Foundation, Delaware, U.S.A.) was used in Spyder integrated development environment (IDE) version 5.2.2. The entire process involves normalizing images by gain and exposure settings and then converting into radiance followed by reflectance. After that unsharp mask (an image sharpening technique) was applied as enhancement technique to improve visual sharpness and then gamma correction was applied to make the enhanced images appear brighter and visually closer to what our eyes see (MicaSense Incorporated, 2022). The unsharp mask uses linear filter to add a fraction of high-pass filtered input image to the original image that helps to sharpen the original image by filtering out the noise (Allebach, 2005). Similarly, gamma correction is an image enhancement technique that is used to minimize the effect of non-linearity of the imaging sensors thereby making the images appear brighter with enhanced contrast and visually closer (Guo et al., 2004; Ju et al., 2018; Xu et al., 2009).

**YOLOv5**

Since its release in June of 2020, YOLOv5 (Jocher et al., 2021b) has become a popular algorithm for object detection in CV applications. YOLOv5 was originally released in four different variants: YOLOv5s, YOLOv5m, YOLOv5l and YOLOv5x, with the subscript based on the network depth and number of parameters used. Here, *s*, *m*, *l* and *x* represent small, medium, large and extra-large variants of the YOLOv5 network respectively. In this study, we used the scaled down version of YOLOv5m which was 308 layers deep and contained 21,060,447 parameters. The YOLOv5 network comes with pretrained weights from training on the Common Objects in Context (COCO) (Lin et al., 2014) dataset, which consists of 80 different classes, such that YOLOv5 is pretrained to detect 80 different classes. In this study the network was customized to detect a single class, "*vc*", for VC plants. The overall architecture of YOLOv5m contains 25 nodes (also known as modules), which are named from *model 0* to *model 24* (fig. 5). Modules 0 to 9 form the backbone network, while modules 10 to 23 represent the neck network, and the *model 24* forms the detection layer. This last module consists of three layers to make detections at three different scales.

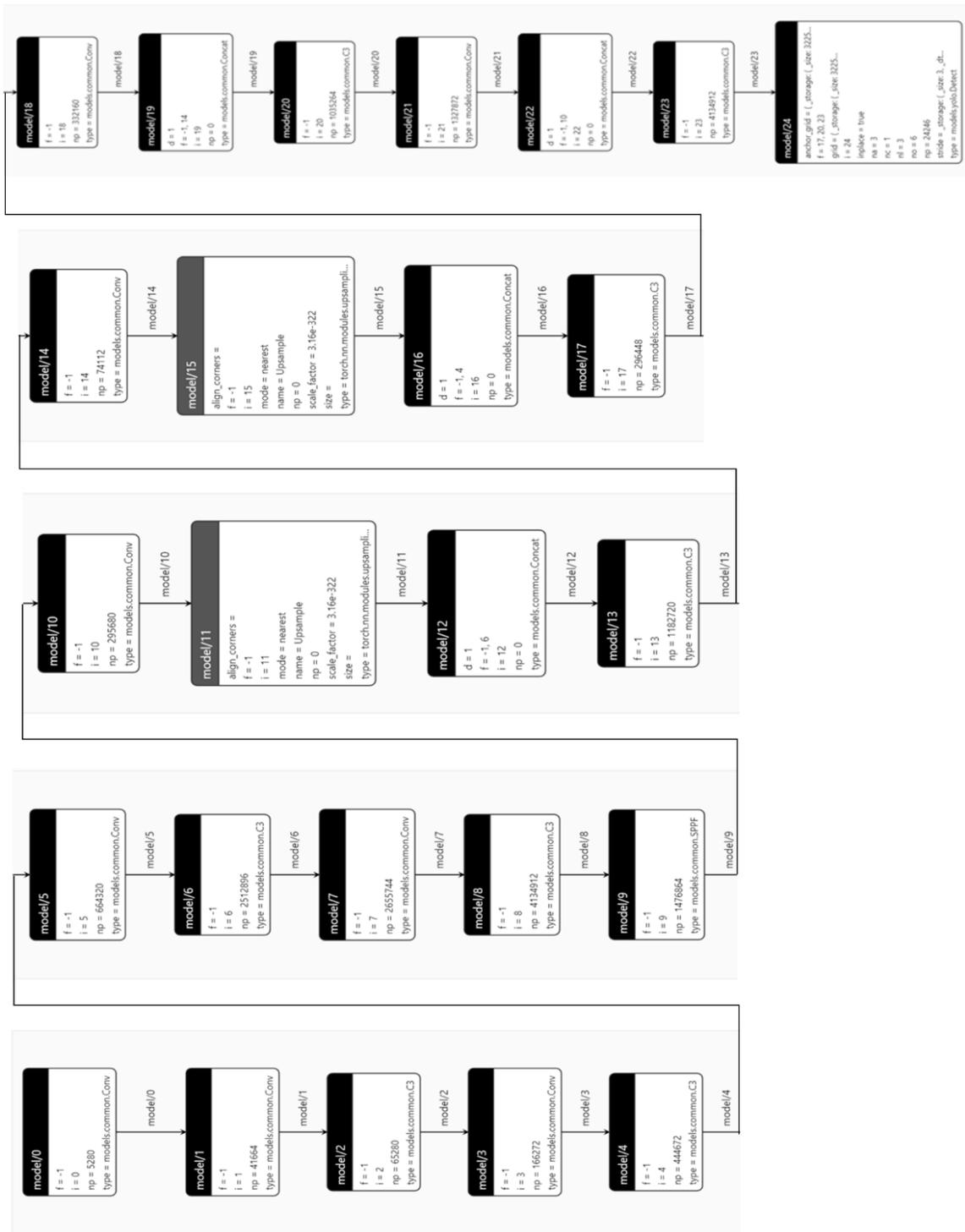

**Figure. 5. YOLOv5m network architecture as generated by Netron (Lutz, 2017) visualization software.**

**IMAGE DATA PREPARATION FOR TRAINING of YOLOv5M**

Among the radiometrically and gamma corrected RGB images, only the images containing VC plants were chosen and then image augmentation was applied using Augmentor Python library (Bloice, 2017). For this, *rotate*, *flip_left_right*, *zoom_random* and *flip_top_bottom* were used with probability values of 0.80, 0.40, 0.60 and 0.80 respectively. These values were chosen so that each time an image was passed through the augmentation pipeline, 80% of the time *rotate* operation was applied, 40% of the time *flip_left_right* was applied, 60% of the time *zoom_random* was applied and 80% of the time *flip_top_bottom* was applied. In this way, we were able to generate a total of 521 RGB images from a total of 34 images (containing at least one VC plants) that were radiometrically and gamma corrected and each of them were of the original size 1207 x 923 pixels. Out of these 417 images (80%) were used for training, 77 (15%) were used for validation and 27 (5%) were used for testing.

**YOLOv5M TRAINING**

The YOLOv5m source code was obtained from the GitHub repository of Ultralytics Inc. (Jocher et al., 2021). The PyTorch framework (Facebook AI Research Lab, Melno Park, CA, U.S.A.) with torch version 1.10.0 and Compute Unified Device Architecture (CUDA) version 11.1.0 (NVIDIA, Santa Clara, CA, U.S.A.) were used to train the YOLOv5m model on Tesla P100-PCIE-16 GB (NVIDIA, Santa Clara, CA, U.S.A.) GPU using the Google Colaboratory (Google LLC, Melno Park, CA, U.S.A.) AI platform. The model was trained with the original hyperparameter values as initial learning rate of 0.01, final learning rate of 0.1, momentum of 0.937, weight decay of 0.0005 and intersection over union (IoU) threshold of 0.20 for a total of 621 iterations in two runs i.e., in the first run it was trained for a total of 500 iterations and the best weights was found, after that in the second run, the training was started from the best weight obtained in the first run for a total of 600 iterations; however the training process had early stopping at the 121$^{st}$ iteration since no improvement could be observed in the 100 iterations past the 21$^{st}$ in the second run. This essentially implied that YOLOv5m had reached the convergence within the 621 total iterations.

**BOUNDING BOX COORDINATE CONVERSION**

The first step involved in this was to extract the top left corner coordinates from the geotagged images of RedEdge-MX camera during the process of radiometric and gamma corrections, image enhancement and RGB band alignments by modifying the original Python script obtained from the GitHub repository of MicaSense (MicaSense Incorporated, 2022). The extracted GPS coordinates of each geotagged image were stored in a comma separated variable (CSV) file. The YOLOv5 *detect.py* Python script was modified to extract central coordinates of each detected bounding boxes that were stored in a separate CSV file. Another Python script was developed to utilize both the CSV files and then convert the pixel-wise BB of the detected VC plants into GPS coordinates. In the Python script, GSD was first converted into meter/pixel and the decimal coordinates were converted into Universal Transverse Mercator (UTM) from which northing and easting values were extracted and then pixel-based central coordinates were converted into UTM based GPS coordinates using UTM zone as 14 during the conversion process ("State of Texas UTM Zones," 2011)

**OPTIMAL FLIGHT PATH WITH ACO ALGORITHM**

ACO algorithm is state-of-the-art for many problems like vehicle routing, open-shop scheduling and sequential ordering problems (Dorigo et al., 2021). It is based on the fact that when ants move along a path from their colony to the food source, they deposit pheromones which evaporates over time. This means ants travelling through the longer paths have less intense pheromone deposit but the pheromone intensity is much higher along the shortest path. In our used case application, VC plant locations are the food source that the artificial ants will find through the shortest possible route. In many past cases, ACO has proven to be widely accepted algorithm for determining optimal flight paths for UAS (Sun et al., 2020; Zhang et al., 2010). Source code to implement ACO was used from the GitHub repository of *fabien-brulport* (Fabien-brulport, 2022). The original source code was modified to generate a CSV file containing GPS coordinates (latitude, longitude) in the order of nodes generated for the optimal route as the output from the ACO algorithm. In our study, we used the values of pheromone weight, heuristic weight, and evaporation rate as 2.01, 1, 0.5 respectively to achieve the shortest distance after trying with different combination of values. Similarly, the values for number of agents i.e., number of artificial ants (i.e., equivalence of number of UAS) and number of iterations were 1 and 100 respectively. Mathematically, at any given time *t*, the probability that ant *k* choses a path from *i* to *j* as explained by Zhang et al. (2010) is given by:

$$P_{ij}^{(k)}(t) = \left\{ \frac{[\tau_{ij}(t)]^\alpha [\eta_{ij}(t)]^\beta}{\sum_{S \subset allowed_k}[\tau_{is}(t)]^\alpha [\eta_{is}(t)]^\beta} \text{ if } j \in allowed_k\text{ ; 0 otherwise} \right\} \quad (1)$$

where, $\alpha$ and $\beta$ are pheromone and heuristic weights respectively. $\eta_{ij}(t)$ and $\tau_{ij}(t)$ are visibility and amount of pheromone between points *i* and *j*. $S \subset allowed_k$ is a set of all the possible points in a path that the ant can choose from. The pheromone amount determines the visibility which in turn determines the probability of choosing a path. Therefore, higher pheromone amount increases the visibility of a particular path between points *i* and *j* and hence the probability of choosing the path increases.

**SPOT-SPRAY UAS SIMULATION**

Once the geographic coordinates were saved in optimal order corresponding to nodes obtained from ACO, the DroneKit-SITL (3D Robotics, Berkeley, CA, U.S.A.) Python package was used to simulate flight paths of UAS with a Python script and monitored on two ground control stations (GCS): Mission Planner version 1.3.76 (Oborne, 2010) and MAVProxy version 1.8.45 (Tridgell and Barker, 2009). The MAVLink (Koubaa et al., 2019; Meier, 2009) protocol was used to communicate between the two GCS and DroneKit-SITL. This communication was accomplished by using transmission control protocol (TCP) for the master port and then port forwarding the output to three user datagram protocols (UDP) ports. The following steps were used to implement the entire process:

i. Two terminal windows were opened from the Anaconda (Anaconda, Inc., Austin, Texas, U.S.A.) environment.

ii. In both the terminal windows, paths were set to the directory where MAVLink was installed.

iii. The following command was used in one of the terminal windows : *dronekit-sitl copter --home=30.534351,-96.431239,0,180 --model=copter* to start DroneKit SITL with the default copter version installed on the system at a GPS location set in *--home*.

iv. In the second terminal window, the following command was executed : *mavproxy -- master tcp:127.0.0.1:5760 --sitl 127.0.0.1:5501 --out udp:127.0.0.1:14550 --out udp:127.0.0.1:14551 --out udp:127.0.0.1:14552* to port forward the master TCP port of MAVProxy to three UDP ports.

v. Mission Planner GCS software was opened and then the mode was set to *guided* by using *Set Mode = Guided* after connecting to UDP port 14551

vi. MAVProxy GCS console opened automatically in which "MAVProxy" in the menu bar was clicked to open "Map' for visualizing the simulated UAS.

vii. Then Spyder IDE was opened from the previously opened Anaconda environment in step i.

viii. We made sure the path was set to *.conda\envs\tensorflow2\Lib\site-packages\pymavlink.*

ix. Then Python script "*vc_spot_spray.py*" was run to simulate spot-spray applications based on the optimal flight path generated by the ACO algorithm.

The Python script in step ix was used to read the CSV file that contained GPS coordinates of all the nodes in order generated by the ACO algorithm which corresponded to the locations of the detected VC plants for spot-spraying. After the UAS simulation was performed successfully, we were also able to upload the CSV file with optimal flight paths on AgroSol version 2.105.0 GCS developed by Hylio (Hylio, Inc., Richmond, Texas, U.S.A.). The AgroSol software provides interface to control the spot-spray UAS that was customized. The flowchart in figure 6 shows the entire workflow starting from data collection to spot-spray UAS simulation.

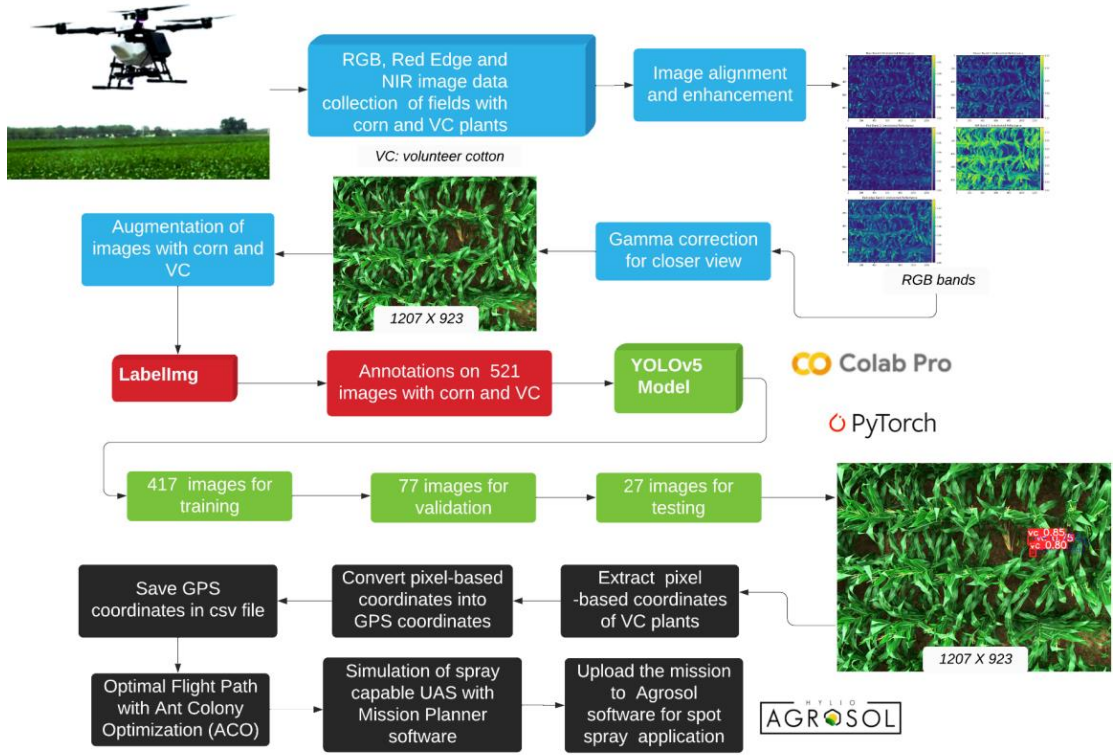

**Figure 6. A flowchart that shows complete workflow representing each step used in this study.**

# RESULTS

## CV ALGORITHM WITH YOLOv5M FOR DETECTING VC PLANTS IN A CORN FIELD ON RADIOMETRICALLY CORRECTED AERIAL IMAGERY

Figure 7 shows graphs for box loss (IoU loss) and objectness loss on both training and validation datasets. Before the 600$^{th}$ iteration, both the losses had converged and no improvement was seen beyond. The lowest value of box loss was found to be 0.028 at 476$^{th}$ iteration while the lowest value for objectness loss i.e., 0.009 was found at iteration number 613 on the training dataset. On the validation dataset, the lowest value of objectness loss was found to be 0.0094 at 394$^{th}$ iteration. Similarly, the lowest value of box loss on the validation dataset was found to be 0.0296 at iteration number 370. From figure 8, graphs of different performance metrices can be seen. It was found that the maximum value of precision reached around 0.98 at 320$^{th}$ iteration while the maximum value for recall was found to be 0.77 at 525$^{th}$ iteration. The most important metric out of all of these is mAP@0.50 whose maximum value i.e., 0.81 reached at iteration number 613. At iteration number 521, the maximum value for mAP@0.50:0.95 was obtained which was found to be 0.33. The precision-recall curve (PRC) resulted in an overall accuracy of nearly 79% and the maximum value of F1-score was found to be 0.76 at nearly 40% confidence level (figs. 9A and 9B). The confusion matrix in figure 9C shows that YOLOv5m was trained enough to classify VC plants with an accuracy of 78% and loss of 22% owing to the background class of corn plants, weeds, soil, etc.

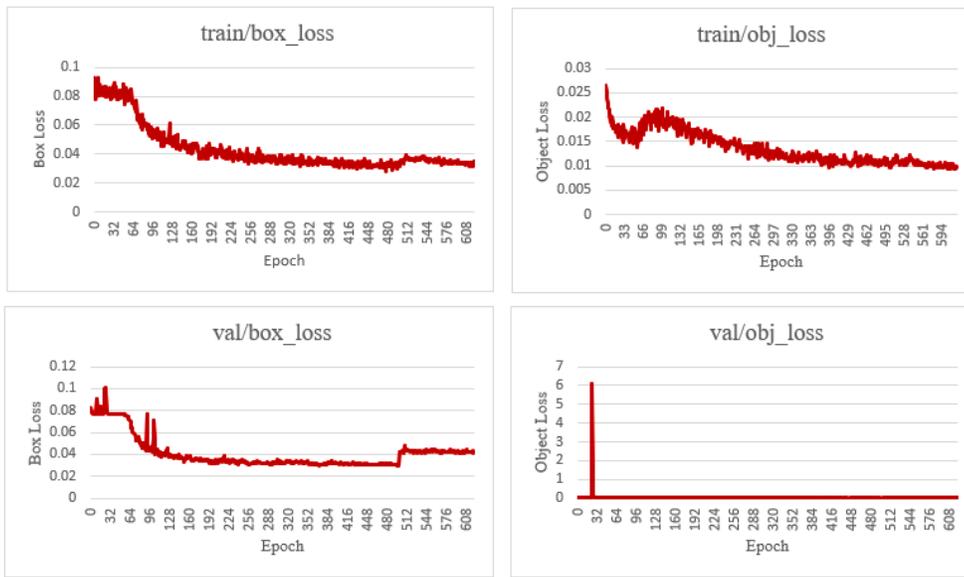

**Figure 7. Different types of losses that were obtained during the training process of YOLOv5m on training and validation datasets.**

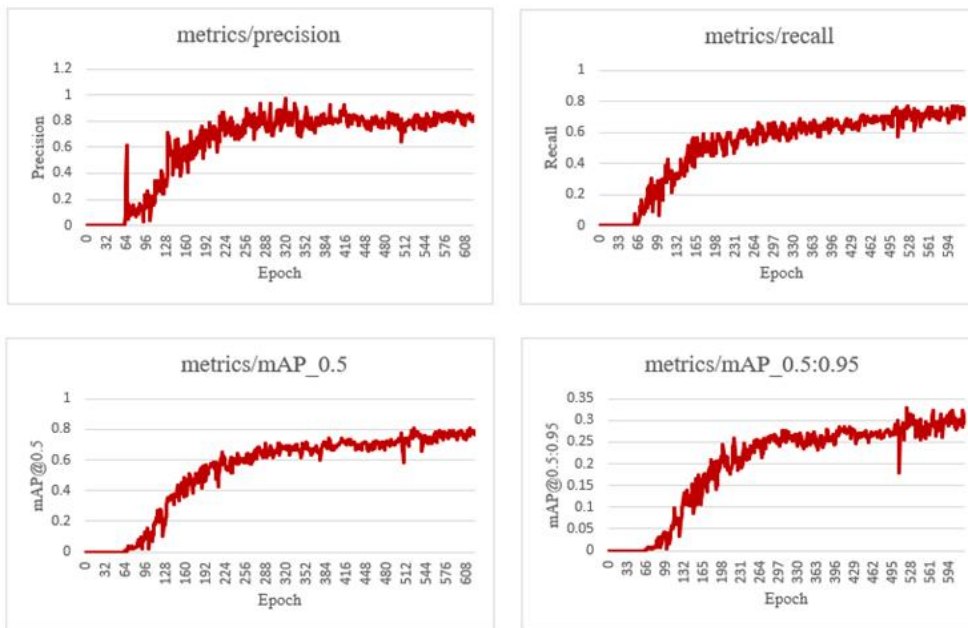

**Figure 8. Different types of performance metrics that were obtained during the training process of YOLOv5m.**

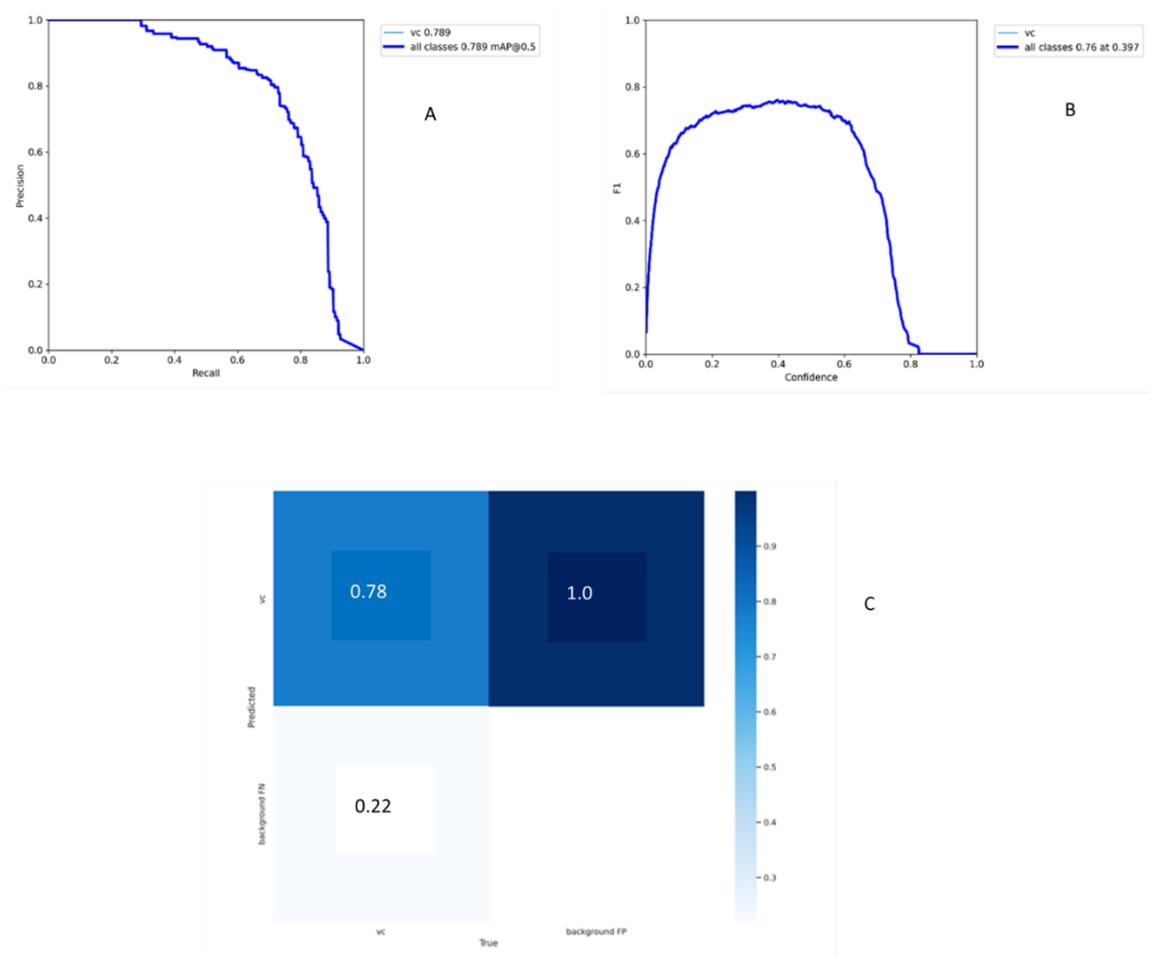

**Figure 9. (A) Precision-Recall plot, (B) F1-score vs confidence score plot and (C) confusion matrix obtained after training YOLOv5m.**

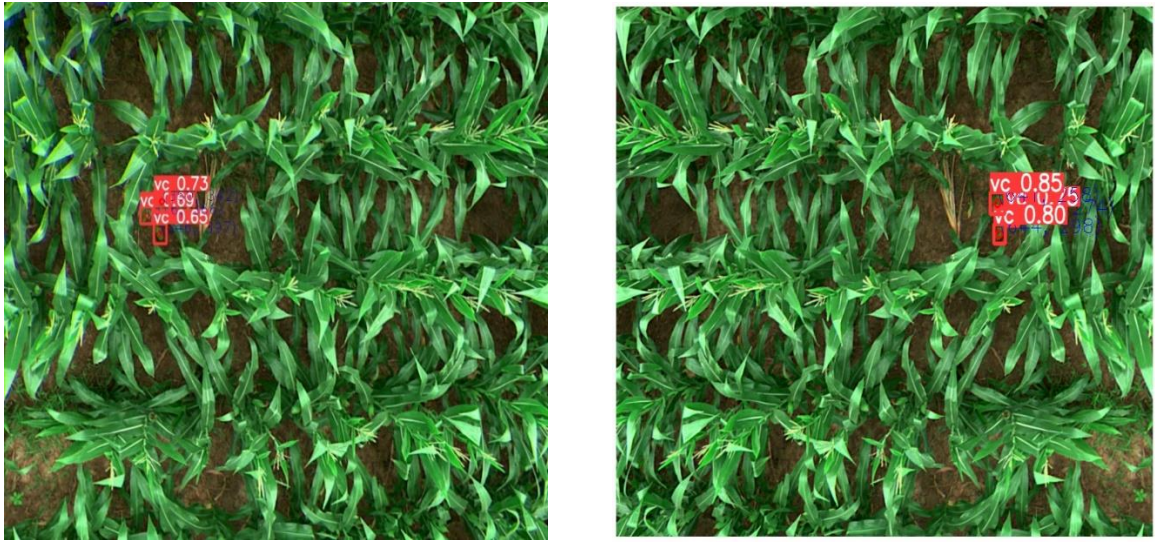

**Figure 10. VC plants detected in the middle of a corn field within the red bounding boxes (BB) by trained YOLOv5m model. The values associated with each BB show how confident the model is about VC plants within it.**

Figure 10 shows some detection results by the trained YOLOv5m model that resulted an average inference speed of 47 frames per second (FPS) on NVIDIA Tesla P100 GPU-16GB. It was later deployed on NVIDIA Jetson TX2 GPU that was mounted on the custom UAS (fig. 2) that resulted on an adjusted average inference speed of 2.535 seconds (~ 0.4 FPS) for images of size 640 x 640 pixels (fig. 11).

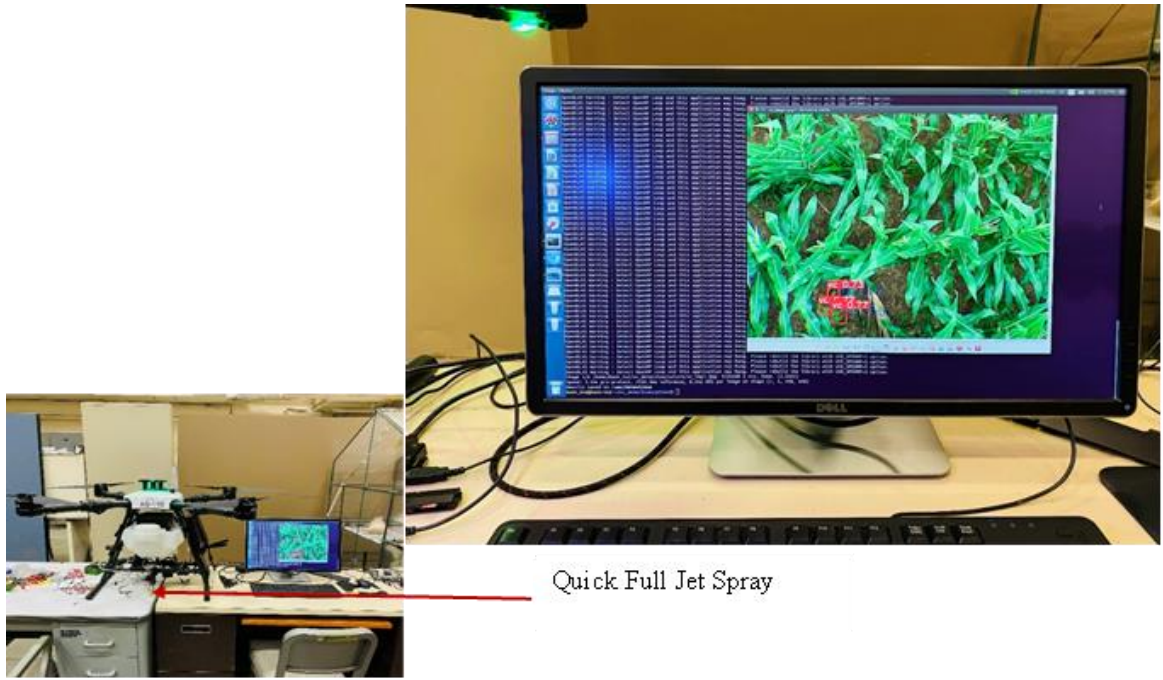

**Figure 11. YOLOv5m detection of VC plants in a corn field by deploying on NVIDIA Jetson TX2 mounted on a custom spot-spray capable UAS.**

## ACO ALGORITHM TO DETERMINE OPTIMAL FLIGHT PATH FOR SPOT-SPRAY APPLICATIONS

Ten random locations of VC plants in the experimental plot were chosen whose corresponding GPS coordinates can be seen on the left-side of table 1. Each location consisted of multiple VC plants (since at each location multiple cotton seeds were planted); however only one plant's GPS location was considered. This represents mimicking real-world scenario in which VC plants usually grow in groups and by considering location of one plant from each group, the entire spot consisting of the group of VC plants can be sprayed.

**Table 1. GPS coordinates of randomly chosen locations of ten VC plants in experimental plot**

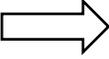

A

| VC | Latitudes | Longitudes |
|---|---|---|
| 1 | 30.5343 | -96.4312 |
| 2 | 30.53431 | -96.4301 |
| 3 | 30.53492 | -96.4303 |
| 4 | 30.53386 | -96.4299 |
| 5 | 30.53531 | -96.4289 |
| 6 | 30.53484 | -96.4299 |
| 7 | 30.53687 | -96.4284 |
| 8 | 30.53537 | -96.4289 |
| 9 | 30.5359 | -96.4283 |
| 10 | 30.53321 | -96.4298 |

ACO

B

| Nodes | Latitudes | Longitudes |
|---|---|---|
| 1 | 30.53687 | -96.4284 |
| 2 | 30.5359 | -96.4283 |
| 3 | 30.53321 | -96.4298 |
| 4 | 30.53386 | -96.4299 |
| 5 | 30.53431 | -96.4301 |
| 6 | 30.53492 | -96.4303 |
| 7 | 30.5343 | -96.4312 |
| 8 | 30.53484 | -96.4299 |
| 9 | 30.53537 | -96.4289 |
| 10 | 30.53531 | -96.4289 |
| 11 | 30.53687 | -96.4284 |

In this case, the first location was also considered to be the home location of spot-spray UAS at which the flight begins and ends. The right side of table 1 shows the generated nodes in the order determined by the ACO algorithm. Nodes 1 and 11 are the same location from which the flight begins and at which it ends. The optimal path generated for spot-spray application can be seen on a webpage-based output (fig. 12) that was generated by *streamlit* Python package (Streamlit Inc., San Francisco, CA, U.S.A.). The bottom graph shows number of iterations on the x-axis and the optimal distance covered on the y-axis (in Kilometers). The total distance covered by the spot-spray capable UAS along the generated path was found to be 674.17 meters (0.67 Kilometers/0.42 miles).

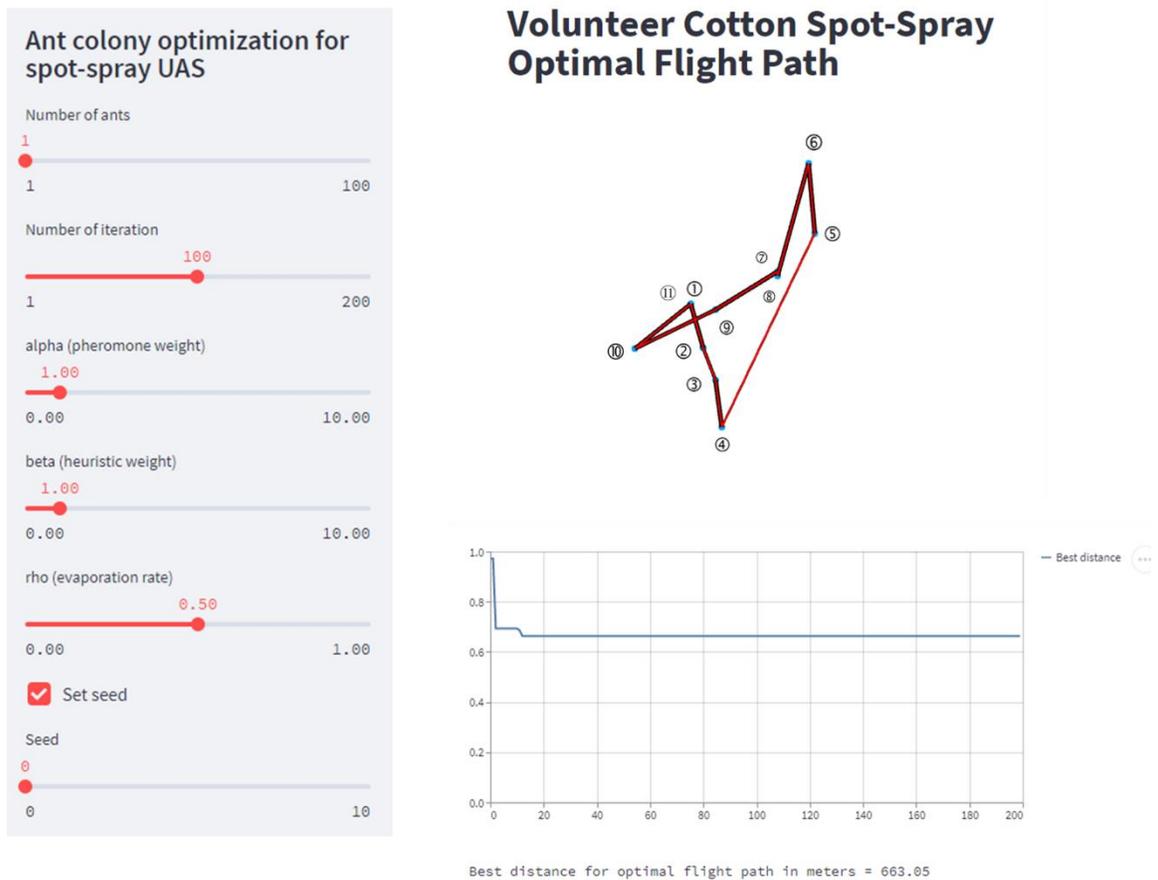

**Figure 12. Optimal flight path generated by ACO algorithms and output shown by *streamlit* Python package on a webpage.**

## SPOT-SPRAY UAS SIMULATION ON MAVPROXY AND MISSION PLANNER BASED ON THE OPTIMAL FLIGHT PATH GENERATED BY ACO ALGORITHM

Once the optimal flight path was obtained (table 1-B) by the ACO algorithm, the DroneKit-SITL was used to simulate the spot-spray UAS by following the steps described under the section, *spot-spray UAS simulation*, on both the GCS-MAVProxy and Mission Planner as shown in figures 13-A, B and C.

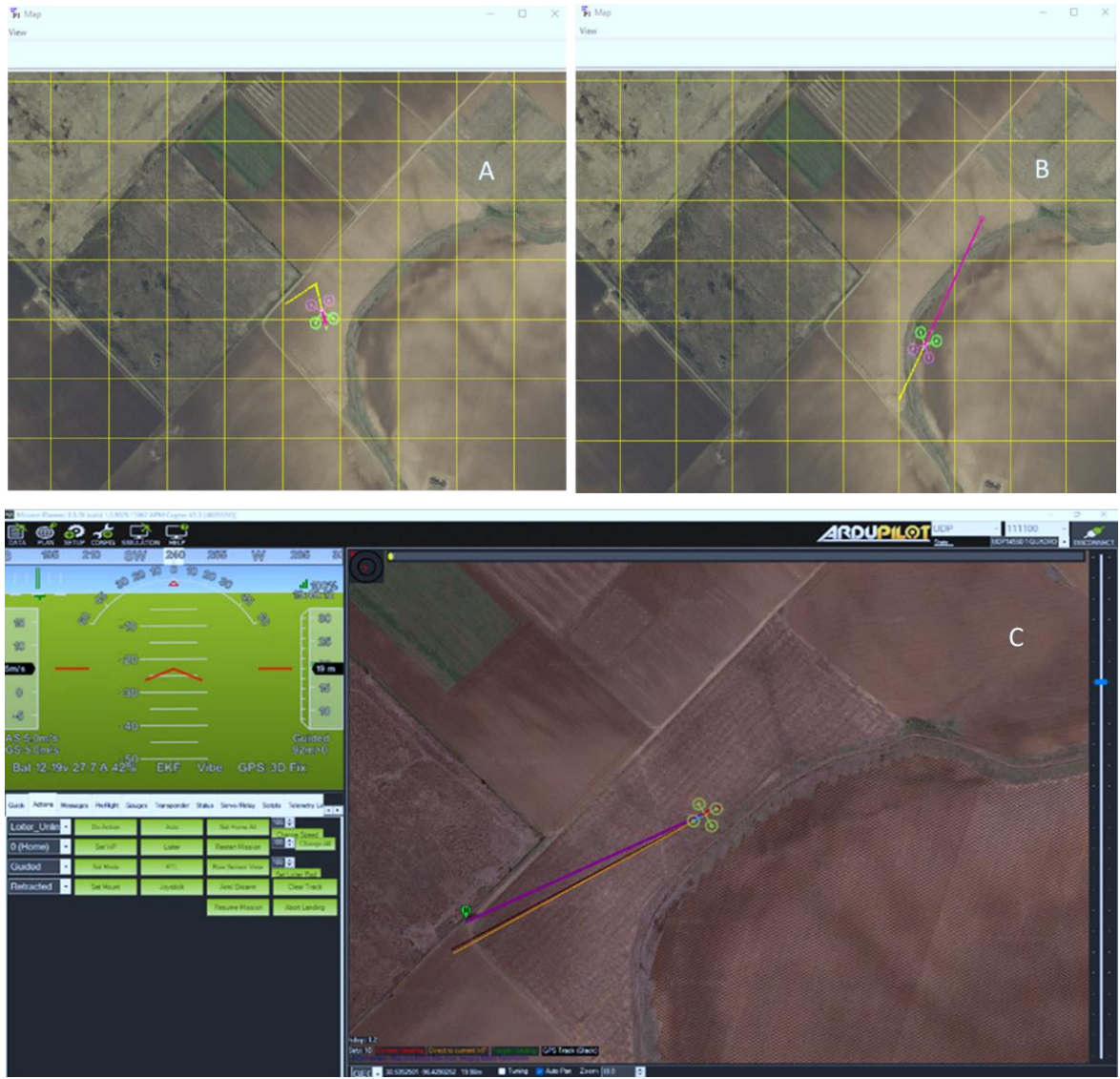

**Figure 13. Spot-spray UAS simulation on MAVProxy (A,B) and Mission Planner (C) GCS. Image *A* shows the simulated UAS flying from node 1 to 2 while image *B* shows from node 4 to 5. Image *C* shows the simulated UAS flying from node 8 to 9.**

**SPOT-SPRAY MISSION ON AGROSOL GCS**

AgroSol (Hylio Inc., Richmond, Texas, U.S.A.) GCS software allows to upload the CSV file containing GPS coordinates of nodes generated by the ACO algorithm and then different settings for spot-spray applications can be used for real life spray applications. (fig. 14). Once the CSV file containing the spot locations was uploaded, AgroSol generated a spot group as seen on the left side of figure 14. It then allows to fill in the values for different parameters like spray altitude, spray volume, etc. The spray mission can be uploaded to the UAS flight controller and then the UAS can go to each of the spot locations and precisely spot-spray at each of them.

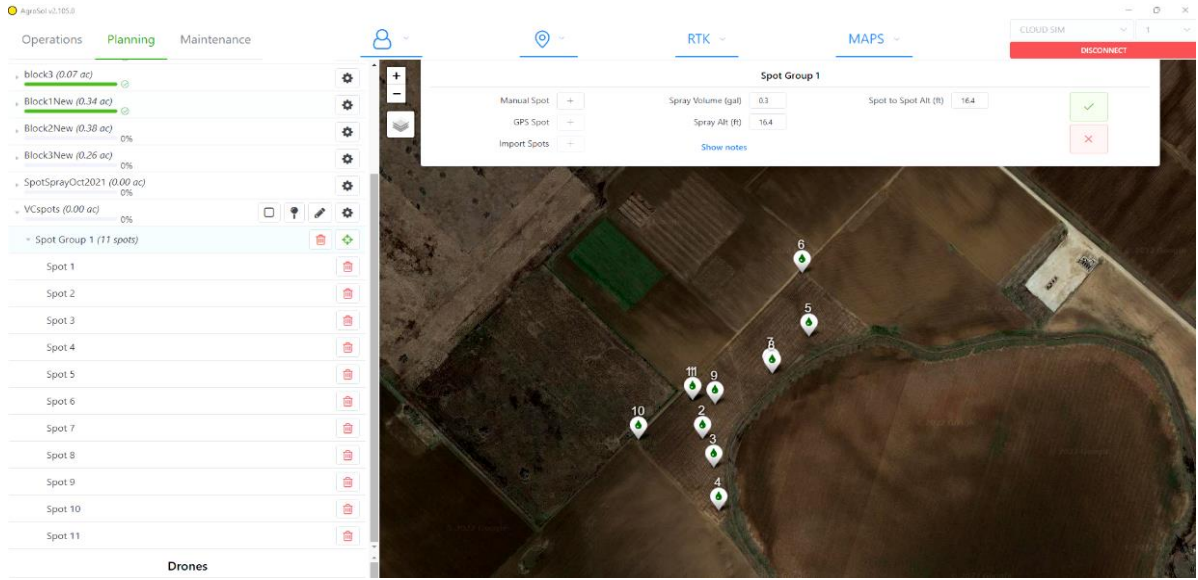

**Figure 14. Spot-spray nodes generated by Agrosol software after uploading the CSV file containing nodes generated by ACO algorithm.**

# DISCUSSION

## VC DETECTION WITH YOLOv5M

The potential application of multispectral remote sensing imagery for VC detection in other crops was mentioned by Zhang et al. (2012) when they were developing methods of discriminating cotton plants from other crops based on spectral reflectance properties. Similarly, in our previous study (Yadav et al., 2019), we were able to show that UAS based multispectral remote sensing imagery can be used to detect VC plants growing in corn fields. In another study by Westbrook et al. (2016), they were able to detect early growth stage VC plants with aerial remote sensing RGB imagery. Study reported in this paper takes motivation and recommendations from all the previous studies as it makes use of multispectral remote sensing imagery collected by UAS. However, the difference lies in the use of CV algorithm as opposed to conventional image processing techniques like discriminant and principal component analysis (PCA), maximum likelihood classification, linear spectral unmixing, etc. In the study of Zhang et al. (2012), they were able to discriminate cotton plants from other crops like corn, soybean and sorghum with 100% accuracy; however, the method used handheld spectroradiometer or the ones mounted on a tractor very close to crop canopies. This means that their method was limited by two major factors: the first one is that it cannot be applied to larger areas of field and the second one is that it is a time-consuming process that is not suitable for near real-time detection of VC plants. Our previous study (Yadav et al., 2019) used remote sensing multispectral imagery and some classical machine learning techniques to make the process of VC detection semi-automatic; it however, couldn't produce classification accuracy greater than 70%.

In this study, we were able to develop a CV algorithm with YOLOv5m to detect VC plants (before pinhead square growth stage) growing in a corn field at tassel (VT) growth stage using multispectral remote sensing imagery. In another study (chapter 2), we were able to detect VC plants with more than 90% accuracy in early growth corn field (V3 vegetative growth stage) with uncalibrated RGB imagery. YOLOv3 was used which resulted in nearly 29% higher classification accuracy than the previous approach (Yadav et al., 2019) but the trained model was less robust to illumination conditions and environmental factors as the images were not radiometrically corrected. Moreover, the relative height difference between VC and corn plants was less compared to the study reported in this paper.

In our current and latest approach reported herein, we were able to develop a CV algorithm based on YOLOv5m which is more robust to illumination conditions and environmental factors as it was trained on radiometrically corrected imagery. By training the YOLOv5m model on radiometrically corrected imagery we were able to generate a more reliable model because it was trained on more accurate data (Mamaghani and Salvaggio, 2019). With the current approach, we were able to classify VC plants at 11.43% higher accuracy than the previous method (Yadav et al., 2019) as seen in figure 9C. It is also noteworthy to mention that similar detection accuracy was obtained in a corn field when they were at VT stage unlike the V3 stage reported in the other study. Another important advantage of using YOLOv5m in our current study is that we were able to obtain an average inference speed of 11 FPS (fig. 11) as opposed to 2 FPS with YOLOv3 when deployed on the Pascal GPU of NVIDIA Jetson TX2 development board. This showed that the trained YOLOv5m can be used for near real-time detection of VC plants in corn fields.

The overarching goal of this study was to speed up the management aspect of Texas Boll Weevil Eradication Program (TBWEP) by simultaneously reducing the chemical costs. Therefore, apart from developing a system for near real-time detection of VC plants, it is also required to develop a detection algorithm that results in minimal false negatives (i.e., maximum recall). This way, we can minimize the possibility of missing VC plants in the middle of corn fields that can potentially act as hosts for the boll weevil pests. In other words, for our used case scenario, our algorithm can be tolerant to false positives but not to false negatives as the damages incurred by boll weevil pests are much more than the extra costs incurred due to spraying some undesired locations because of false positives. For this reason, the model obtained at $525^{th}$ iteration can be used (fig. 8). However, if high precision is required (like in the case of saving chemical costs), then the model trained at iteration $322^{nd}$ can be used which results in precision as high as 98% (fig. 8). The overall performance of the model is represented by mAP@50 whose maximum value was found to be 81% at $613^{th}$ iteration (fig. 8). The performance of the trained model can be improved by adjusting the confidence threshold value mentioned by Yan et al. (2021). Apart from these, we had generated 521 images from limited number of datasets i.e., 34 images by using image augmentation techniques. In this way the trained model reported in this study can be regarded as relatively more generalizable (Gan et al., 2021)

**ACO ALGORITHM FOR OPTIMAL FLIGHT PATH AND SPOT-SPRAY**

UAS flight parameters such as aerial speed, flight altitude, yaw angle, etc. were assumed to be fixed and the optimal path generated in figure 12 was based on equal weights assigned to both pheromones and heuristic parameters. These assumptions were similar to the ones made by Ma et al. (2007). The only constraint used in our implementation was distance between two GPS locations. However, in other studies, parameters like yaw angle were used as one of the constraints (Zhang et al., 2010). The ACO algorithm being stochastic in nature generates sufficiently good solution based on randomly generated variables but not globally optimal solution (Bianchi et al., 2009). Therefore, the optimal path shown in figure 3.14 may not be the global optimal but simply based on the randomly generated variables used in our study. The parameter values used for the ACO in this study may not be the best ones but were tested with reasonable combinations. From past studies, it was found that ACO is highly sensitive to the evaporation rate which is considered to be equivalent of learning rate; therefore choosing the right value of this parameter is crucial (Ebadinezhad, 2020; Ojha et al., 2015). Ojha et al. (2015) found that the performance of ACO decreased beyond the 0.5 value for evaporation rate. Hence, we chose to use 0.5 as the evaporation rate in our case (fig. 12). A total of 11 nodes were generated from the 10 VC plant locations as seen in Table 3.1-left and 3.1-right. Nodes 1 and 11 are the same location from which the UAS begins and ends it journey traversing through the optimal path generated (figs. 12, 13 and 14).

## CONCLUSIONS AND FUTURE WORK
### CONCLUSIONS

In this paper, we were able to show that CV algorithm based on YOLOV5m can be used to detect VC plants (before reaching the pinhead squaring phase) growing in a corn field at tassel growth stage using UAS remote sensing multispectral imagery. We were able to show this with improved classification and detection accuracies when compared to the previous methods. In this paper, we were able to demonstrate that low resolution imaging sensor can be used with some preprocessing algorithms (radiometric and gamma corrections) to detect VC plants with a model that is more robust to illumination and environmental conditions.

Apart from the improved detection accuracy and developing a more robust CV algorithm, we were able to demonstrate that the trained YOLOV5m model can be used for near real-time detection by deploying it on a computing platform mounted on a sprayer UAS. The trained YOLOV5m model was able to detect VC plants of full-scale (1207 x 923 pixels) images at an adjusted average inference speed of nearly 0.4 FPS on a Pascal GPU of Jetson TX2 (NVIDIA, Santa Clara, CA, U.S.A.) development board.

In the end, we were able to convert pixel-wise BB central coordinates of detected VC plants into GPS coordinates to use them for generating optimal flight path with ACO algorithm for spot-spray applications. Overall, we were able to develop a CV algorithm for near real-time detection of VC plants in a corn field and spot-spray applications with a customized spray capable

UAS. Therefore, through this research work, we were able to develop a system that has the potential to speed up the mitigation efforts of TBWEP at a reduced management cost.

**FUTURE WORK**

Our future work will involve improving the recall value to minimize false negatives so the number of missing VC plants for spot-spray application can be minimized. We will also compare results of detection accuracies at different growth phases of corn plants i.e., at different relative heights between corn and VC plants. Another aspect of future work will involve testing spray deposition efficacy from different spray heights, wind speeds, etc. as explained in a test study of Martin et al. (2019).

**ACKNOWLEDGEMENTS**

Sincere thanks is extended to all the reviewers, farm manager (Stephen P. Labar) and student assistants (Roy Graves, Madison Hodges, Sam Pyka, Reese Rusk, Raul Sebastian, JT Womack, John Marshall, Katelyn Meszaros, Lane Fisher, and Reagan Smith) involved during field work. This material was made possible, in part, by Cooperative Agreement AP20PPQS&T00C046 from the USDA's Animal and Plant Health Inspection Service (APHIS).